\documentstyle[12pt]{article}

\newcommand{\nn}{\nonumber\\}\newcommand{\p}[1]{(\ref{#1})}
\begin{document}
\renewcommand{\thefootnote}{\fnsymbol{footnote}}
\thispagestyle{empty}
{\hfill   JINR E2-95-250}\vspace{2.5cm} \\

\begin{center}
{\large\bf On the structure of the $N=4$ Supersymmetric \\
Quantum Mechanics in $D=2$ and $D=3$} \vspace{1.5cm} \\
V. Berezovoj\footnote{BITNET: KFTI\%KFTI.KHARKOV.UA@RELAY.USSR.EU.NET}
\vspace{1cm} \\
{\it Institute of Physics and Technology}\\
{\it 310108, Kharkov, Ukraine}\\
and \vspace{1cm}\\
A. Pashnev\footnote{BITNET: PASHNEV@THSUN1.JINR.DUBNA.SU}\vspace{1cm}\\
{\it JINR--Laboratory of Theoretical Physics} \\
{\it Dubna, Head Post Office, P.O.Box 79, 101 000 Moscow, Russia}
\vspace{1.5cm} \\
{\bf Abstract}
\end{center}
The superfield formulation of two - dimensional $N=4$
Extended Supersymmetric Quantum Mechanics (SQM) is described.
It is shown that corresponding
classical Lagrangian describes the motion in the conformally flat
metric with additional potential term.
The Bose and Fermi sectors of two- and three-dimensional $N=4$
SQM are analyzed.
The structure of the
quantum Hamiltonians is such, that the usual Shr\"{o}dinger
equation in the flat space arises after some unitary transformation,
demonstrating the effect of transmutation of the coupling constant and the
energy of the initial model in some special cases.
\begin{center}
{\it Submitted to ``Classical and Quantum Gravity"}
\end{center}
\vfill
\setcounter{page}0
\renewcommand{\thefootnote}{\arabic{footnote}}
\setcounter{footnote}0
\newpage
\section{Introduction}

The simplest way to construct the classical Lagrangian of the SQM in
$ D $ dimensions is to
consider superfields $\Phi^i(\tau,\eta^a) , \; i = 1, 2, ... D $ in the
superspace $\tau,\eta^a \; a =1, 2, ... N$
with one bosonic and $  N $ grassmann coordinates. The first
component of the superfields are  the usual bosonic coordinates $X^i$ ,
the next ones $\psi^{ia}$ are grassmann coordinates.
So, the classical Lagrangian of the Supersymmetric Quantum Mechanics (SQM)
describes the evolution of bosonic and additional grassmann degrees
of freedom, which after quantization become generators of the Clifford
algebra. This fact naturally leads to the matrix realization of the
Hamiltonian and Supercharges of SQM \cite{W}.

The dimensionality of
such realization depends on the number of grassmann variables and in the
case of scalar superfields $\Phi^i$ it is $2^{[\frac{DN}{2}]}$. So, it
rapidly growths for extended supersymmetry and, for example,
 takes the value $d=64$ for
$D=3, \; N=4 $ case. The way out of this difficulty is to use more
complicated representations of extended supersymmetry \cite{AP} -  \cite{BP}.
The main idea
The simplest of them is the chiral superfield, which contains one complex
bosonic and $\frac{N}{2} $ complex grassmann fields. The Lagrangian for
such superfield naturally describes the two - dimensional SQM.
The main idea of the reduction of the number of
grassmann degrees of freedom for other values of $D$ is to use superfields
$\Phi^i$ which transform nontrivially under the isomorphism algebra
of the extended supersymmetry algebra
\begin{equation}
\{ Q^a , Q^b \} = i\delta^{ab}\frac{\partial}{\partial\tau}.
\end{equation}
If $N=4$ the isomorphism algebra is $SO(4)= SO(3)\times SO(3)$ and
index $i$ can describe three-dimensional vector representation of one of
the $SO(3)$  algebras which plays simultaneously the role of space
rotations algebra. The dimensionality of the matrix representation
is $d = 4$. The Hamiltonian and Supercharges for such
description of three-dimensional $N=4 $ SQM were obtained in
\cite{BP}. In the present paper we describe the two-dimensional
$N=4$ SQM with the help of the chiral superfield. Along with two
dynamic bosonic coordinates such superfield contains four dynamic
grassmann coordinates and dimensionality of the matrix representation
for Hamiltonian and Supercharges is four again.

In the third and forth sections we analyze Bose and Fermi sectors of the
two-dimensio- nal and three-dimensional SQM in such minimal superfield
approach.
Despite of the complicated structure of the gravitation-like
interaction with additional nontrivial potential energy,
the resulting stationary
equations after some unitary transformation become the usual
stationary Shr\"{o}dinger equations, demonstrating the coupling
constant - energy transmutation in some cases.

\setcounter{equation}0\section{Two - dimensional SQM}
In this section we construct the two - dimensional $N=4$ SQM in the frames
of the superfield approach.
The $N=4$ supersymmetry algebra can be written in terms of two
complex supercharges $Q_a, \;a=1,2$ and hamiltonian $H$
\begin{equation} \label{n4}
\{ Q_a ,{\overline Q}^b \} =\delta_a^b H ,\;
[ H,Q_a ] = [ H,{\overline Q}^a ] =0,\;{\overline Q}^a=Q_a^\star.
\end{equation}
Its automorphism group is $SO(4)= SU(2) \times SU(2)$ and $Q_a$ transforms
as a spinor of one of the $SU(2)$ groups.

The chiral superfield
\footnote{Our conventions for spinors are as follows:
$ {\overline \theta}_a \equiv {(\theta^a)^\star} ,\theta_a \equiv \theta^b
\varepsilon_{ba}, \theta^a =\varepsilon^{ab} \theta_b, {\overline \theta}^a
\equiv \varepsilon^{ab} {\overline \theta}_b, {\overline \theta}_a =
{\overline \theta}^b \varepsilon_{ba} ,{\overline \theta}^a =-(\theta_a)^\star,
(\theta \theta ) \equiv \theta^a \theta_a =-2 \theta^1 \theta^2 ,
( {\overline \theta} {\overline \theta} ) \equiv {\overline \theta}_a
{\overline \theta}^a  = ( \theta \theta )^\star .$}
\begin{equation}
\Phi(\tau,\theta,\overline{\theta})=Z(\tau)+\theta^a\chi_a(\tau)+
\frac{i}{2} \overline{\theta}\theta\dot{Z}(\tau)+
\theta\theta F(\tau)-\frac{i}{4}\theta\theta \overline{\theta}_a\dot{\chi}^a-
\frac{1}{16}\overline{\theta}\overline{\theta}\theta\theta\ddot{Z}(\tau)
\end{equation}
in the superspace with one bosonic coordinate $\tau$ and two complex
fermionic coordinates $\theta^a$ behaves as a scalar
under the supersymmetry transformations
\begin{equation}
\delta\theta^a = \epsilon^a,\; \delta\overline{\theta}_a = \overline{
\epsilon}_a,\;\delta\tau = \frac{i}{2}(\epsilon^a\overline{\theta}_a
+\overline{\epsilon}_a\theta^a)
\end{equation}
and satisfies the chirality conditions $\overline{D}^a\Phi=0$, where
\begin{equation}
\overline{D}^a = \frac{\partial}{\partial\overline{\theta}_a}-
\frac{i}{2}\theta^a\frac{\partial}{\partial\tau}
\end{equation}
are the supersymmetric covariant derivatives.

The most general action for the superfield $\Phi$
\begin{equation}\label{action}
S = \frac{1}{2}\int d \tau d^2 \theta d^2 \overline{\theta} V(\Phi,
\overline{\Phi})+\int d \tau d^2 \theta R(\Phi)+\int d \tau
d^2 \overline{\theta} \overline{R}(\overline{\Phi}).
\end{equation}
contains one real function $V(\Phi,\overline{\Phi})$ and one chiral
function $R(\Phi)$.
After the integration over $\theta$ and $\overline{\theta}$ we find the
component Lagrangian, in which fields $F$ and $\overline{F}$ are
auxiliary and they can be dropped with the help of their equations of
motion.

Finally, the component Lagrangian takes form:
\begin{eqnarray}
L&=&\frac{1}{2} W \dot{z} \dot{\overline{z}}+
i(\psi^a \dot{\overline{\psi}}_a)
-\frac{i}{W}(\dot{z}\frac{\partial W}{\partial z}-\dot{\overline{z}}
\frac{\partial W}{\partial \overline{z}})\psi\overline{\psi}-
\frac{\partial}{\partial {z}}\left(\frac{{U}}{W}\right)
\psi \psi-
\frac{\partial}{\partial \overline{z}}\left(\frac{\overline{U}}{W}\right)
\overline\psi \overline\psi+
\nn
&&+\frac{2}{W^2}(\frac{\partial^2 W}{\partial z \partial \overline{z}}
-\frac{1}{W}\frac{\partial W}{\partial z}\frac{\partial W}{\partial
\overline{z}})\psi\psi\overline{\psi}\overline{\psi}-\frac{U
\overline{U}}{2W},
\end{eqnarray}
where we have introduced
\begin{equation}
W(z ,\overline{z})=\frac{{\partial}^2 V(z, \overline{z})}{\partial z
\partial \overline{z}},\;\;U(z)=\frac{\partial R(z)}{\partial z},\;\;
 \psi=\frac{\sqrt{W}}{2}\chi.
\end{equation}

The quantization procedure takes into account the dependence of the
Lagrangian on the conformally flat metric $g_{ik}\dot{z}^i\dot{z}^k=
W(z, \overline{z}) \dot{z} \dot{\overline{z}}$ \cite{BP} and leads to the
following quantum Hamiltonian and Supercharges:
\begin{eqnarray}
\hat H&=& \frac{2}{W}{\cal P}\overline{{\cal P}} +\frac{2i}{W^2}
(\frac{\partial W}{\partial z}\overline{{\cal P}}  -
\frac{\partial W}{\partial \overline{z}}{\cal P})(\frac{1}{2}-
\overline{\psi}\psi)+\\
&&\frac{4}{W^2}(\frac{3}{2W}\frac{\partial W}{\partial z}
\frac{\partial W}{\partial \overline{z}}-\frac{{\partial}^2 W}{
\partial z \partial \overline{z}})(\frac{1}{2}-\overline{\psi}\psi)^2
+\frac{\partial}{\partial {z}}\left(\frac{{U}}{W}\right)
\psi \psi+
\frac{\partial}{\partial \overline{z}}\left(\frac{\overline{U}}{W}\right)
\overline\psi \overline\psi+\frac{U\overline{U}}{2W},\nn
\hat Q^{a}&=& (\pi\psi^{a}+i\frac{\overline{U}}{2}
\overline{\psi}^{a})\frac{2}{\sqrt{W}}, \\
\hat{\overline{Q}}_{a}&=&\frac{2}{\sqrt{W}}
 (\overline{\psi}_{a}\overline{\pi}+
i\frac{U}{2}\psi_{a}),
\end{eqnarray}
where
\begin{equation}
\pi={\cal P}+\frac{i}{W}\frac{\partial W}{\partial z}
(\psi\overline{\psi}-\frac{3}{2}) ,
\overline{\pi}=\overline{{\cal P}}-\frac{i}{W}\frac{\partial W}{\partial
\overline{z}}
(\psi\overline{\psi}-\frac{1}{2}) ,
\end{equation}
 and ${\cal P}=-i\frac{
\partial}{\partial z},$ $ \overline{{\cal P}}=-i\frac{\partial}{\partial
\overline{z}}  $.
Grassmann variables $\overline{\psi}_b$ and $\psi^a$ satisfy the following
commutation  relations
\begin{equation}
\{\psi^{a},\overline{\psi}_{b}\}=\frac{1}{2}\delta^a_b,\;a,b=1,2\; .
\end{equation}
and can be considered as a
 creation and annihilation operators.
The general quantum state can be written as the vector in the
corresponding Fock space
\begin{equation}
|\Phi(z, \overline{z}) \rangle = \phi^1 (z, \overline{z})|0\rangle
 +\chi^a(z, \overline{z}) \overline \psi_a |0\rangle +
\phi^2 (z, \overline{z}) \overline \psi^a \overline \psi_a |0\rangle ,
\end{equation}
where $|0\rangle$ is the vacuum of the Fock space: $ \psi^a |0\rangle =0$.
The Hamiltonian $\hat H$
is hermitian and hermiticity properties of supercharges
${\hat{\overline{Q^a}}}={\hat Q_a}^{+} $ are fulfilled with respect
to the scalar product
\begin{equation}
( \Phi_1,\Phi_2 ) = \int d^2 z W(z, \overline{z}) \langle \Phi_1(
z, \overline{z})|\Phi_2(z, \overline{z})
\rangle .
\end{equation}
The only case when the scalar product \p{sp1} coincides with the usual
scalar product is $W(z,\overline{z}) =const$.

The normalizable solutions of the stationary Schr\"{o}dinger equation
\begin{equation}
\hat H |\Phi \rangle =E |\Phi \rangle
\end{equation}
describe the physical states.
In Bose sector this equation is non diagonal
and leads to two connected equations for wavefunctions
$\phi^A(z, \overline{z})$
\begin{eqnarray}                   \label{bose}
&&\{ \frac{2}{W}{\cal P}\overline{{\cal P}} +\frac{i}{W^2}
(\frac{\partial W}{\partial z}\overline{{\cal P}}  -
\frac{\partial W}{\partial \overline{z}}{\cal P})+
\frac{1}{W^2}(\frac{3}{2W}\frac{\partial W}{\partial z}
\frac{\partial W}{\partial \overline{z}}-\frac{{\partial}^2 W}{
\partial z \partial \overline{z}})+
\frac{U\overline{U}}{2W}\}\phi^1(z, \overline{z})+\nn
&&\frac{\partial}{\partial \overline{z}}\left(\frac{\overline{U}}{W}
\right)
\phi^2(z, \overline{z})= E \phi^1(z, \overline{z})\\ \label{bose1}
&&\{ \frac{2}{W}{\cal P}\overline{{\cal P}} -\frac{i}{W^2}
(\frac{\partial W}{\partial z}\overline{{\cal P}}  -
\frac{\partial W}{\partial \overline{z}}{\cal P})+
\frac{1}{W^2}(\frac{3}{2W}\frac{\partial W}{\partial z}
\frac{\partial W}{\partial \overline{z}}-\frac{{\partial}^2 W}{
\partial z \partial \overline{z}})+
\frac{U\overline{U}}{2W}\}\phi^1(z, \overline{z})+\nn
&&\frac{\partial}{\partial {z}}\left(\frac{{U}}{W}\right)
\phi^2(z, \overline{z})= E \phi^2(z, \overline{z})
\end{eqnarray}
It means that true physical states in the case of nonvanishing chiral
terms in the action \p{action} are
superpositions of the bosonic states with different fermionic number $(0$
and $2)$.
After the unitary transformation
\begin{equation}
\hat{H} =W^{-1/2}\hat{\tilde{H}}W^{1/2}, \;\;\;
\phi^A =W^{-1/2}\tilde{\phi^A}
\end{equation}
the system \p{bose}-\p{bose1} takes more simple form
\begin{eqnarray}
\{2{\cal P}\frac{1}{W}\overline{{\cal P}} +\frac{U\overline{U}}
{2W}\}\tilde{\phi}^1(z,\overline{z})
+\frac{\partial}{\partial \overline{z}}\left(\frac{\overline{U}}{W}
\right)
\tilde{\phi}^2(z,\overline{z})&=&
E \tilde{\phi}^1(z,\overline{z})  \\
\{2\overline{{\cal P}}\frac{1}{W}{\cal P} +\frac{\alpha\overline{\alpha}}
{8W}\}\tilde{\phi}^2(z,\overline{z})+
\frac{\partial}{\partial {z}}\left(\frac{{U}}{W}\right)
\tilde{\phi}^1(z,\overline{z})&=&
E \tilde{\phi}^2(z,\overline{z}).
\end{eqnarray}
There is no need to solve this system. The simplest way to find the
solution of this system is to find the solution of the system
in the sector with fermionic number $1$ and to apply one of the
supercharges $Q_a$ to it. As a consequence of the supersymmetry
algebra \p{n4} it will be the solution in the bosonic sector.

In the Fermi sector the Hamiltonian $\hat H$ is diagonal and
equations for the spinors $ \chi^a(z, \overline{z})$ are very simple:
\begin{equation}
\frac{2}{W}({\cal P}{\overline{\cal P}}+\frac{U\overline{U}}
{4})\chi^a(z, \overline{z})=E\chi^a(z, \overline{z}).
\end{equation}
These equations are just the stationary zero energy Shr\"{o}dinger
equations
\begin{equation}
(2{\cal P}{\overline{\cal P}}+\tilde{U}(z, \overline{z}))
\chi^a(z, \overline{z})=0
\end{equation}
with the potential
\begin{equation}\label{ge}
\tilde{U}(z,\overline{z})=
\frac{U(z,\overline{z})\overline{U(z,\overline{z})}}{2}-
EW(z,\overline{z}).
\end{equation}
The whole potential $\tilde{U}$ is combined from two functions
$U(z)$ and $W(z,\overline{z})$. Some particular choices of this
functions are interesting. The simplest of them $W(z,\overline{z})=1$
leads to the standard Shr\"{o}dinger equation with the potential
\begin{equation}
\tilde{U}(z,\overline{z})=
\frac{U(z,\overline{z})\overline{U(z,\overline{z})}}{4}
\end{equation}
and the energy $E$. The opposite situation takes place when
$U(z)=\alpha=const$ and $W(z,\overline{z})$ is arbitrary.
The energy $E$ in the corresponding Shr\"{o}dinger equation
\begin{equation}
(2{\cal P}{\overline{\cal P}}-EW(z, \overline{z}))
\chi^a(z, \overline{z})=
-\frac{1}{2}\alpha\overline{\alpha}
\chi^a(z, \overline{z})
\end{equation}
plays the role of the coupling constant.
In turn,
the coupling constant $\alpha$ , namely its function ${\cal E}=
-\frac{1}{2}\alpha\overline{\alpha}$,
plays the role of the energy. This situation demonstrates the effect
of coupling constant - energy transmutation. As we will see later,
only this situation takes place in three dimensions.

In the general situation of arbitrary $U(z)$ and $W(z,\overline{z})$
the energy $E$ plays the role of the coupling constant and the
physical spectrum is determined by the existence of the normalizable
solutions of the equation \p{ge} at the definite values $E_n$ of the
parameter $E$.

\setcounter{equation}0\section{Hamiltonian and supercharges in $D=3$}
The following expressions for Hamiltonian and supercharges for three -
dimensional SQM have been obtained in \cite{BP}:
\begin{eqnarray}
\hat{H}&=&-\frac{1}{2}\frac{1}{W(x)}{\partial}_i^2-\frac{1}{4}\frac{
\partial_iW(X)}{W^2(x)}\partial_i+\frac{1}{2}\frac{\alpha^2}{W(x)}-
\frac{3}{8}\frac{\partial^2_iW(x)}{W^2(x)}+\frac{15}{32}\frac{(\partial_i
W(x))^2}{W^3(x)}+ \nn
 & & {}\nn
 & &+(\overline \psi_a\psi^a-\frac{1}{2})^2
\left(\frac{\partial_i^2W(X)}{W^2(x)}-\frac{3}{2}\frac{(\partial_iW(x))^2}{
W^3(x)}\right)+ \\
 & & {}\nn
 & & +i\epsilon_{ikl}\psi\sigma_i\overline\psi\frac{
\partial_kW(x)}{W^2(x)}\partial_l-\alpha\psi\sigma_i\overline \psi
\frac{\partial_iW(x)}{W^2(x)},\nn
 & & {}\nn
\hat Q_a&=& \left[(\sigma_i)_a^b\left(p_i
-i(\overline \psi_c\psi^c-\frac{1}{2})\partial_i\ln W(x)\right)
+i\alpha\delta_a^b\right]\frac{\overline \psi_b}{\sqrt{W(x)}},\\
 & & \nn
 {\hat{\overline{Q^a}}} &=&
\frac{\overline \psi^b}{\sqrt{W(x)}}
\left[(\sigma_i)^a_b\left(p_i
+i(\overline \psi_c\psi^c-\frac{1}{2})\partial_i\ln W(x)\right)
-i\alpha\delta_a^b\right],
\end{eqnarray}
where $i,k,l$ are three-dimensional vector indices and $a,b = 1,2$ -
spinor indices as in the two - dimensional case.
Again the  operators $\overline \psi_b$ and $\psi^a$  are creation and
annihilation operators
and the general quantum state can be written as the $x$-depending vector
in the corresponding Fock space
\begin{equation}
|\Phi(x) \rangle = \phi^1 (x)|0\rangle  +\chi^a(x) \overline \psi_a |0\rangle +
\phi^2 (x) \overline \psi^a \overline \psi_a |0\rangle ,
\end{equation}
In 3 dimensions the Hamiltonian $\hat H$ and operators
\begin{equation}
p_i=-i\partial_i-\frac{3}{4}i\partial_iW(x)
\end{equation}
are Hermitian and ${\hat{\overline{Q^a}}}={\hat Q_a}^{+} $ with respect
to the scalar product
\begin{equation}\label{sp}
( \Phi_1,\Phi_2 ) = \int d^3 x W^{3/2}(x) \langle \Phi_1(x)|\Phi_2(x)
\rangle ,
\end{equation}
which contains the measure $W^{3/2}$ in contrast to 2 - dimensional case.

All of the operators (2.1)-(2.3) are completely determined in terms of the
function $W(x)$ (connected with the superpotential $V(x)$:
$W(x)=\partial_i\partial_iV(x)$) and one additional parameter $\alpha$,
which also characterizes the parity violation.
The sum of the angular
momentum and spin operators
\begin{equation}\label{j}
\hat J_i =\epsilon_{ikl}x_kp_l+\psi^a(\sigma_i)_a^b\overline \psi_b
\end{equation}
is conserved operator, describing the total momentum of the system.
The eigenvalues of the operator \p{j} are integer for bosonic states
with wavefunctions $\phi^A(x),\; A=1,2,\; $ and half-integer for fermionic
states with grassmann spinor wavefunctions $\chi^a(x)$.

The normalizable solutions of the stationary Schr\"{o}dinger equation
\begin{equation}
\hat H |\Phi \rangle =E |\Phi \rangle
\end{equation}
describe the physical states. In Bose sector this equation is diagonal
and leads to two identical equations
\begin{equation}
\hat H_B \phi^A(x) = E \phi^A(x)
\end{equation}
with\footnote{to avoid misunderstandings
 we denote the derivative which
acts on everything to the right by means of $\hat{\partial}_i$}
\begin{equation}
\hat H_B = -\frac{1}{2}g^{-\frac{1}{2}}\hat{\partial}_i g^{ij}(x)g^
{\frac{1}{2}}\partial_j +\frac{\alpha^2}{2W}-\frac{1}{16}R,
\end{equation}
where
\begin{equation}
g_{ij}(x) =W(x) \delta_{ij}
\end{equation}
is the metric tensor and
\begin{equation}
R = 2\left ( \frac{\partial_i\partial_iW(x)}{W^2(x)}-\frac{3}{4}
\frac{(\partial_iW(x))^2}{W^3(x)}\right )
\end{equation}
- the scalar curvature of the corresponding three dimensional space.
Note the equality of the coefficient at $R$ to that, calculated in
\cite{CK}.
In the Fermi sector the Hamiltonian $\hat H$ is non diagonal and the
equation for the spinor $ \chi^a(x)$ is as follows:
\begin{eqnarray}
\left\{  -\frac{1}{2}g^{-\frac{1}{2}}\hat{\partial}_ig^{ij}(x)g^{\frac{1}{2}}
\partial_j +\frac{\alpha^2}{2W}-\frac{3}{8}\left ( \frac{\partial_i
\partial_iW(x)}{W^2(x)}-\frac{5}{4}
\frac{(\partial_iW(x))^2}{W^3(x)}\right )\right\} \chi^a(x)-& & \nn
 { }& & \\
-\frac{1}{2}\left \{ i\epsilon_{ikl}\frac{\partial_kW(x)}{W^2(x)}\partial_l-
\alpha\frac{\partial_iW(x)}{W^2(x)}\right \}(\sigma_i)^a_b\chi^b(x) &=&
\chi^a(x).\nonumber
\end{eqnarray}
The first of nondiagonal terms in this equation is proportional to the
spin-orbit coupling and the second one leads to the parity violation.

Both the bosonic and fermionic equations can be written in more simple
form with the help of transformation
\begin{equation}          \label{ct}
\hat{H} = W^{-\frac{1}{4}}(x)\hat{\tilde H} W^{\frac{1}{4}}(x),\;
\phi^A(x) = W^{-\frac{1}{4}}(x)\tilde{\phi}^A(x),\;
\chi^a(x)= W^{-\frac{1}{4}}(x)\tilde{\chi}^a(x).
\end{equation}
The transformed equations are:
\begin{eqnarray}\label{be}
\left\{-\frac{1}{2}\partial_i^2-EW(x)\right\}\tilde{\phi}^A(x)&=&-\frac{1}
{2}\alpha^2\tilde\phi^A(x),\\
\left\{-\frac{1}{2}\partial_i^2-EW(x)+\frac{1}{2}W^{\frac{1}{2}}
\partial_i^2 W^{-\frac{1}{2}}\right\}\tilde{\chi}^a(x)-& & \\
-\frac{1}{2}\left\{ i\epsilon_{ikl}\partial_k{\ln W(x)}\partial_l-
\alpha\partial_i{\ln W(x)}\right\}(\sigma_i)^a_b\tilde{\chi}^b(x)&=&
-\frac{1}{2}\alpha^2\tilde\chi^a(x). \nonumber
\end{eqnarray}
The bosonic ones are just the stationary Shr\"{o}dinger equations
with the potential $U(x)=-EW(x)$.
The energy $E$ plays the role of the coupling constant and
the coupling constant $\alpha$ , namely its function ${\cal E}=
-\frac{1}{2}\alpha^2$, plays the role of the energy.
It means, that the effect of coupling constant - energy metamorphosis
takes place in three-dimensional SQM as well.

In spite of the
fact that the bosonic Schr\"{o}dinger equation \p{be} is written in the
flat three dimensional space, the scalar product for wavefunctions
$\tilde{\phi}^A(x)$ contains the function $W(x)$:
\begin{equation}\label{sp1}
(\phi^A , \phi^B )= \delta^{AB}\int d^3 x W(x) \phi^{A*}(x)\phi^B(x).
\end{equation}
This relation is the consequence of the relations \p{sp}-\p{ct}.

\setcounter{equation}0
\section{Example}
As an illustration  we consider in three dimensions the case
$W(x) =\frac{\kappa}{r},\;r=\sqrt{x_i^2}$.
The scalar product is
\begin{equation}
( \phi_1,\phi_2 ) = \int d^3 x \frac{\kappa}{r} \phi_1^{*}(x)\phi_2(x).
\end{equation}
Taking the standard form for the wavefunction
\begin{equation}
\phi(x)=\frac{1}{r}\sum\limits_{l,m}^{}u_{nl}(r) Y_{lm}(\theta,\phi),
\end{equation}
we find the equation for $u_{nl}(r)$:
\begin{equation}\label{ueq}
u_{nl}^{''}(r)+\left\{ \frac{2\kappa E}{r}-\frac{l(l+1)}{r^2}-\alpha^2
\right\} u_{nl}(r)=0.
\end{equation}
with normalization condition
\begin{equation}
(u_{n'l'},u_{nl})=\kappa \int \frac{dr}{r} u_{n'l'}^{*}(r)u_{nl}(r)=
\delta_{n'n}\delta_{l'l}.
\end{equation}
The normalizable solutions of the equation \p{ueq} are
\begin{equation}
u_{nl}(r)=C_{nl} \; r^l\; e^{-\alpha r}{}_1F_1(l+1-n,2l+2;2\alpha r)
\end{equation}
with constant $C_{nl}$.
The energy spectrum is given by the following relation
\begin{equation}\label{e}
E_{nl}=\frac{\alpha}{\kappa}(n+l+1)
\end{equation}
The solutions of the corresponding equation in the Fermi-sector
for $W(x)=\frac{\kappa}{r}$
\begin{eqnarray}          \label{f}
\left\{-\frac{1}{2}\partial_i^2-\frac{E\kappa}{r}+\frac{3}{8}\frac{1}{r^2}
\right\}\tilde{\chi}^a(x)+& & \\
+\frac{1}{2r^2}\left\{ i\epsilon_{ikl}x_k \partial_l-
\alpha x_i \right\}(\sigma_i)^a_b\tilde{\chi}^b(x)&=&
-\frac{1}{2}\alpha^2\tilde\chi^a(x). \nonumber
\end{eqnarray}
can be obtained from the solutions of
bosonic equation with the help of supersymmetry transformations. The
energy spectrum of fermionic equation \p{f} is also given by the formula
\p{e}.\vspace{0.5cm} \\

\noindent {\bf Acknowledgments.}We would like to thank D.V. Volkov
and E.A. Ivanov for useful discussions and comments on the subject.
This investigation has been supported in part by the Russian Foundation of
Fundamental Research, grant 93-02-03821, the International Science
Foundation, grants M9T300 and UA6000  and INTAS, grant 93-127

\end{document}